\documentclass[aps,prl,preprint,groupedaddress,amssymb,amsmath,floatfix]{revtex4-2} 
\usepackage{xcolor}

\usepackage[colorlinks=true, allcolors=blue]{hyperref}
\usepackage{graphicx}
\usepackage{todonotes}
\usepackage{acronym}
\graphicspath{ {./figures/} }
\usepackage{subcaption}
\usepackage{amsmath}
\usepackage[LGRgreek]{mathastext}
\usepackage{lineno}
\usepackage[english]{babel}

\begin{document}

\title{2D Material Exciton-Polariton Transport\\on 2D Photonic Crystals}
\author{Xin Xie$^{1,2}$}
\author{Qiuyang Li$^{1}$}
\author{Chenxi Liu$^{1,3}$}
\author{Yuze Liu$^{4}$}
\author{Chulwon Lee$^{1}$}
\author{Kai Sun$^{1}$}
\author{Hui Deng$^{1,4*}$}

\affiliation{}
\affiliation{$^1$Department of Physics, University of Michigan, Ann Arbor, Michigan 48109, United States}
\affiliation{$^2$Michigan Institute for Data Science, University of Michigan, Ann Arbor, Michigan 48109, United States}
\affiliation{$^3$Department of Nuclear Engineering, University of Michigan, Ann Arbor, Michigan 48109, United States}
\affiliation{$^4$Department of Electrical Engineering and Computer Science, University of Michigan, Ann Arbor, Michigan 48109, United States}
\affiliation{$^*$dengh@umich.edu}

\begin{abstract}
Transport of elementary excitations is a fundamental property of 2D semiconductors, important for wide-ranging emergent phenomena and device applications. 
While exciton transport reported in 2D materials barely exceeds 1-2~$\mu$m, coherent coupling of excitons with photons to form polaritons allows not only greatly enhanced transport length, but also the potential to leverage photonic mode engineering for novel transport properties. However, conventional vertical cavity or waveguide polaritons are difficult to tune or integrate into photonic circuits. 
Here, we report the transport of transition-metal dichalcogenide polaritons in slab 2D photonic crystals that are highly versatile for tuning, mode-engineering and integration. We show an order-of-magnitude enhancement of the transport length compared to that of bare excitons. We further show the dependence of transport on the polariton dispersion and population dynamics, which we control by varying the photonic crystal design and pumping intensity. 
Stimulated relaxation observed in the system suggests the potential for forming superfluid polaritons with frictionless transport.
These results demonstrate the 2D photonic crystal polariton system as a versatile platform to enhance and manipulate energy transport for novel photonic technologies.
\end{abstract}

\maketitle

\section{Introduction}
Atomically thin transition-metal dichalcogenides (TMD) have emerged as a new class of semiconductors, which exhibits rich collective phenomena and promises ultra-compact room-temperature exciton and polariton devices with novel functionalities
\cite{Wang_Electronics_2012a,Schaibley_Valleytronics_2016a,Mueller_Exciton_2018,Jauregui_Electrical_2019,Shimazaki_Strongly_2020,Wilson_Excitons_2021,Regan_Emerging_2022,Mak_Semiconductor_2022,Fowler-Gerace_Transport_2024}.
Of critical importance to both basic studies and the applications of TMD semiconductors is long-range transport of the excitations that is easy to tune and control. There have been intense efforts to understand the spatio-temporal dynamics of excitons~\cite{Wang_Ultrafast_2012,Kulig_Exciton_2018,Perea-Causin_Exciton_2019,Uddin_Neutral_2020,Steinhoff_Microscopic_2021,Wagner_Nonclassical_2021,Rosati_Nonequilibrium_2021} and to control exciton transport in TMD monolayers, including using strain~\cite{Li_Optoelectronic_2015,Rosati_Straindependent_2020,Thompson_Anisotropic_2022,Lee_Driftdominant_2022,Datta_Spatiotemporally_2022}, electric field~\cite{Unuchek_Roomtemperature_2018,Unuchek_Valleypolarized_2019,Tagarelli_Electrical_2023}, local dielectric environment~\cite{Hao_Controlling_2020,Li_Dielectric_2021} and moir\'{e} patterns~\cite{Yuan_Twistangledependent_2020}. 
In these works, excitons in TMD monolayers generally exhibit short-range transport of $<2~\mu$m, comparable to the coherence length of a stationary single exciton, due to relatively strong scattering and slow diffusion. 
A promising strategy to significantly enhance the transport length is to coherently couple excitons to photon modes, forming matter-light hybrid quasi-particles of polaritons~\cite{weisbuch_observation_1992,Liu_Strong_2015}.
Transport over tens of microns has been reported in planar structures~\cite{Hu_Imaging_2017,barachati_interacting_2018,wurdack_motional_2021,Guo_Boosting_2022}, due to the high group velocity and low scattering of the photon components. 
However, these polaritons have limited flexibility for integration, tuning or mode engineering. For example, vertical Fabry-P\`erot cavity polaritons~\cite{wurdack_motional_2021} are difficult to tune or integrate with other device components on the chip. Waveguide-like polaritons~\cite{Hu_Imaging_2017,barachati_interacting_2018,Guo_Boosting_2022} have approximately linear dispersions with limited tuning and are incompatible with the formation of polariton condensates for superfluid transport~\cite{Amo_Superfluidity_2009}.

On the other hand, 2D slab photonic crystals (PhCs) have been well established as extremely versatile for mode engineering and integration~\cite{Joannopoulos_Photonic_2011}. 
Since 2D materials are agnostic to substrate materials or interfaces, they have the unique advantage of direct integration with slab PhCs, forming polaritons~\cite{Zhang_Photoniccrystal_2018} that inherit the design and integration flexibility of PhCs. 
For example, parabolic dispersions of different curvatures are expected to facilitate the formation of polariton superfluids. Topological Chern bands with edge states have been predicted, which would allow edge states that are robust against defects or backscattering~\cite{he_polaritonic_2023}. Experimentally, polariton edge states in spin Hall PhCs have been demonstrated~\cite{Liu_Generation_2020, Li_Experimental_2021}.
However, polariton transport has not been reported in these 2D PhC systems, and whether the patterned substrate hinders transport remains unknown. 

Here, we report the first study of 2D material polariton transport on 2D PhCs. We demonstrate enhanced transport length of TMD-PhC polaritons by over an order of magnitude compared to excitons on a flat substrate.  
Furthermore, using an array of PhCs of varying parameters, together with the uniformity and large size of our dodecanol-encapsulated monolayer~\cite{Li_Macroscopic_2023}, we perform the first systematic study of the dependence of the transport properties on the cavity and polariton-mode properties. 
We show that the transport properties are controlled by both the polariton dispersion and polariton relaxation, allowing tuning via 2D PhC designs and excitation intensity. Efficient polariton relaxation through bosonic stimulated scattering was also observed.  
These findings present exciting opportunities for controlling the transport of collective excitations in 2D semiconductors for on-chip exciton and polariton devices.

\section{Results}

\begin{figure}
\centering
\includegraphics[width=0.98\linewidth]{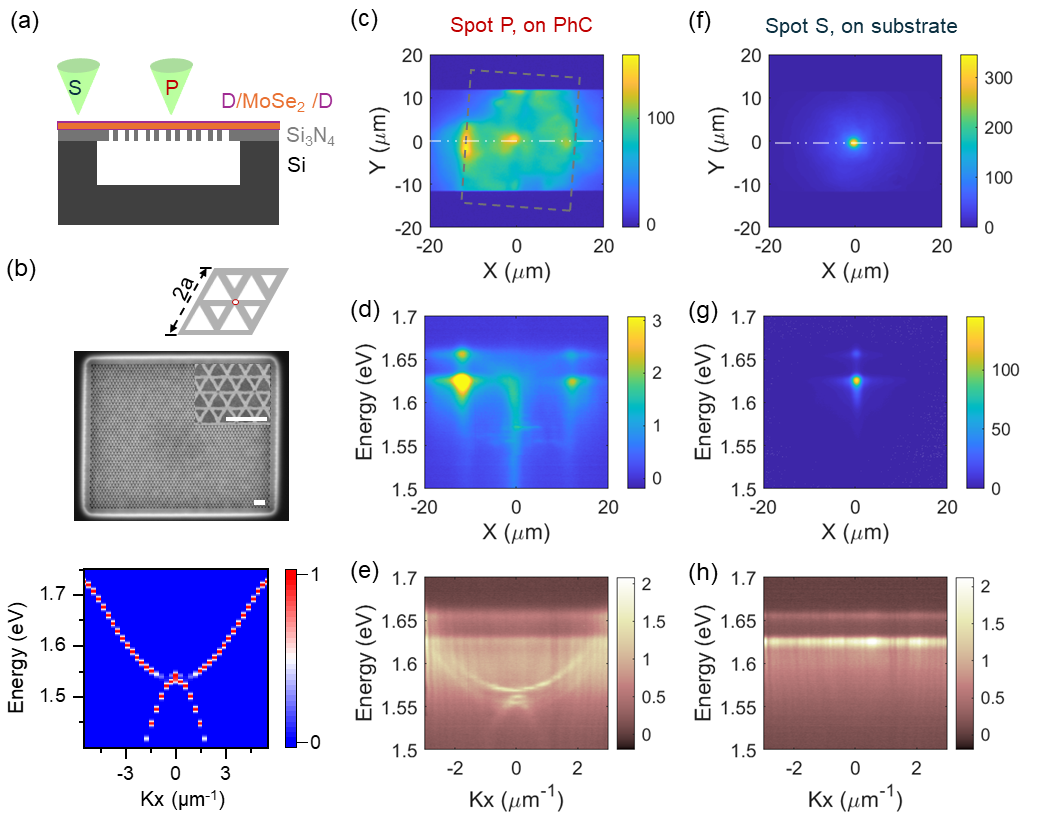}
\caption{\label{f1} Properties of PhCs and comparison of transport features of PhC-polaritons vs excitons. \textbf{(a)} Schematic of a device, composed of a suspended 2D PhC with a large-area dodecanol-encapsulated MoSe$_2$ (D/MoSe$_2$/D) monolayer on top. \textbf{(b)} Schematic of a unit cell of the PhC (top), the SEM image of a fabricated PhC (middle) with inset providing an enlarged view, and the band structure with period $a = 380$~nm (bottom), calculated using Ansys Lumerical FDTD software. The PhC is composed of a honeycomb lattice of two inverted triangular airholes with a period of $a$, along with a triangular lattice of circular airholes (red outline) with a period of $2a$. The scale bars are 1~$\mu$m. \textbf{(c-e)} Measurements of PL of MoSe$_2$ on a PhC. \textbf{(c)} The 2D spatial profile of P. The excitation spot is at (0, 0), the dashed line indicates the boundary of the PhC.  \textbf{(d)} spectra for ($x, 0$) as indicated by the white horizontal dashed lines in (c). \textbf{(e)} Momentum resolved spectra. \textbf{(f-h)} Same measurements as in (c-e) but for PL from MoSe$_2$ on the flat Si$_3$N$_4$ substrate.
}
\end{figure}

The device used in this study consists of arrays of suspended 2D PhCs with a large-area dodecanol-encapsulated MoSe$_2$ monolayer placed on top, as illustrated in Fig. \ref{f1}(a). 
The PhC is made of a Si$_3$N$_4$ film eteched into a honeycomb lattice of triangular and circular holes.
Its unit cell is shown in Fig.~\ref{f1}(b) (top) together with a scanning electron microscopy (SEM) image of a fabricated device (middle). 
The total width of each PhC varies between $15-30~\mu$m. Figure~\ref{f1}(b) (bottom) shows the simulated band structure of a PhC with a period $a = 380 nm$, showing two inverted bands near the exciton energy.
A macroscopic MoSe$_2$ monolayer crystal of several millimeters across is exfoliated by a gold tape and subsequently transferred to the chip and encapsulated by 1-dodecanol molecular layers~\cite{Li_Macroscopic_2023}, covering an array of PhCs with varying lattice parameters. 

We first compare the main features of polariton versus exciton transport through spatially-, spectrally- and momentum-resolved photoluminescence (PL) spectroscopies. The steady-state PL from MoSe$_2$ on the PhC (Spot P, Figs.~\ref{f1}(c-e)) and on the flat Si$_3$N$_4$ substrate (Spot S, Figs.~\ref{f1}(f-h)) show strikingly different features. On the flat substrate, the PL is localized over 1-2~$\mu$m around the excitation spot of about $1~\mu$m in diameter, the PL energy is concentrated at the exciton and trion energies of 1.656~eV and 1.626~eV, and the emission is dispersionless as seen in momentum resolved spectra. In contrast, on the PhC, the PL spatially spreads over the full PhC and is enhanced at the boundary of the PhC, 10s of microns away from the pump spot. Spectrally it spreads to more than 100~meV below the exciton energy (Fig. \ref{f1}(d)), corresponding to the strong polariton dispersion (Fig. \ref{f1}(e)) resulting from the strong PhC dispersion (Fig. \ref{f1}(b) bottom). 
These features show enhanced transport due to the formation of polaritons. They are observed consistently in PhCs of different sizes and with the period of the unit cell varying from $a=260$~nm to $a=440$~nm. 

\begin{figure}
\centering
\includegraphics[width=0.98\linewidth]{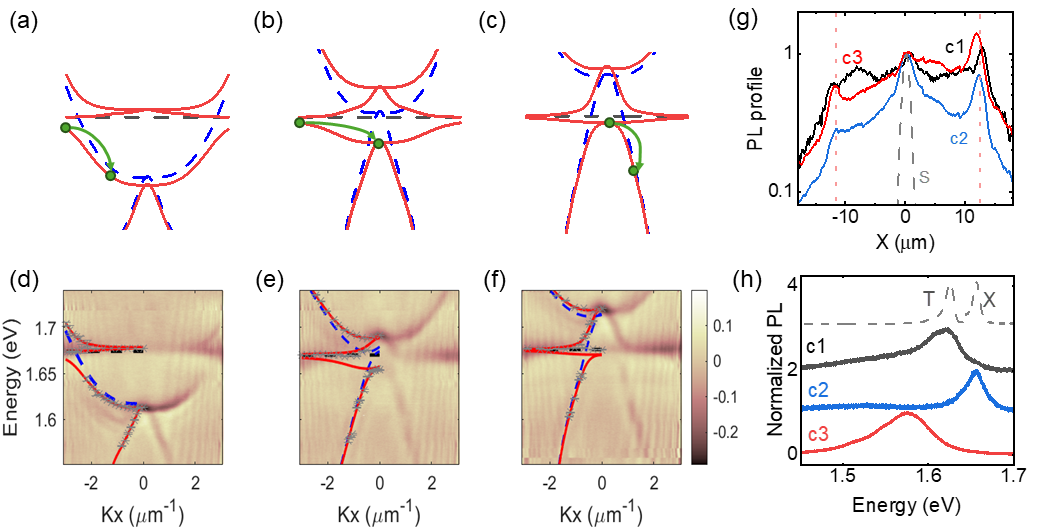}
\caption{\label{f2} Comparison of polaritons with different bands structures. \textbf{(a-c)} Schematics of three different types of lower polariton bands obtained with different photon-exciton detuning, with correspondingly different relaxation processes depicted by the green arrows. The three PhCs have decreasing lattice constant $a$: (a) c1, $a$=375~nm, (b) c2, $a=365$~nm and (c) c3, $a=350$~nm, leading to blueshifts of the photonic bands. \textbf{(d-f)} Momentum-resolved reflection contrast spectra of PhC devices corresponding to c1, c2 and c3 in (a-c). In (a-f), the solid red lines represent the polariton bands, the black and blue dashed lines represent the uncoupled exciton and photon bands, respectively. The gray crosses in (d-f) are fitted results. \textbf{(g)} Spatial profiles of PL integrated over 1.64 to 1.655 eV, from the three PhC devices c1 (black), c2 (blue) and c3 (red) and from spot S over the substrate (gray dashed). The PL intensity is normalized by the value at $X=0~\mu$m. The pink vertical dashed lines indicate the boundary of PhC. \textbf{(h)} PL spectra at the excitation spot from the three PhC devices and spot S. Each curve is normalized to its maximum value.
} 
\end{figure}

Since we have a uniform, macroscopic monolayer simultaneously coupled to an array of PhCs with different lattice parameters, we can further study how mode tuning controls the polariton transport and dynamics.
We compare three different PhCs, PhC c1, c2, and c3, with different types of polariton dispersions, as illustrated in Figs. \ref{f2}(a-c). These PhCs have the same lattice structure, but the period $a$ decreases from 375 nm to 350 nm and the sizes of the triangular holes are scaled accordingly. Consequently, the whole photonic bands blueshift and excitons couple to different bands in the three PhCs. The resulting polaritons have positive, relatively flat, and negative masses, respectively, as shown by their momentum-resolved reflection contrast (RC) spectra in Figs.~\ref{f2}(d-f) and depicted in Figs. \ref{f2}(a-c). The vacuum Rabi splittings are measured to be around 35~meV. More details of the fitting of polariton modes are shown in the Supplementary Materials Fig. S1.

In Figs. \ref{f2}(g) and (h), respectively, we compare the spatial and spectral profiles of the PL from the three PhC devices. Those from the exciton on the flat substrate (Spot S) are also plotted as a reference. 
We first note that the contrast is evident between long-range polariton transport in the PhCs and localized exciton transport on the flat substrate, consistent with the PL images shown in Fig.~\ref{f1}. 
Figure~\ref{f2}(g) shows the spatial profiles of the integrated PL from 1.640 to 1.655 eV. All three PhC devices show a much slower decay of the PL intensities with distance than that of the exciton, with strong emission at the boundary due to enhanced scattering. The decay becomes much faster outside the boundaries of the PhCs, as expected of bare excitons.
Figure~\ref{f2}(h) shows that the spectral profiles of PL at the excitation spot, where broader, redshifted emission is observed in all three PhCs compared to the exciton, as expected from the strong dispersions of polariton modes compared to dispersionless exciton modes. 

Interestingly, the three PhC devices also show clear differences. A longer transport length is observed in devices c1 and c3 compared to c2; at the same time, a very large redshift from the exciton energy by 80~meV is measured in c3, much less in c1, and least in c2. 
%
The different behaviors of the three PhC devices can be understood from their different group velocities and relaxation processes due to their different polariton band structures. Most of the non-resonantly excited carriers first form a reservoir of exciton states where the density of states is very high but group velocity is very small. As they relax into polariton states, as indicated by the green arrows in Figs.~\ref{f2}(a-c), they acquire a high group velocity determined by the polariton dispersion. Both c1 and c3 have relatively steep dispersions and large polariton group velocities at 1.640 to 1.655 eV, while c2 has a relatively flat dispersion and smaller group velocity. Therefore c1 an c3 feature longer transport lengths. 
In c1 and c2, the emission energy is peaked at around the energy of polaritons at $K_x\sim 0$ with positive masses, as polaritons accumulate in these states with lower group velocities while further relaxation to the lower branch is suppressed due to the lower density of states. 
In c3, excitons are at about the same energy of the $K_x\sim 0$ states of the lower branch with a negative mass, therefore relaxation takes place along the lower branch to lower and lower energies without accumulation in any particular state, leading to the observed larger redshift. 

\begin{figure}
\centering
\includegraphics[width=0.98\linewidth]{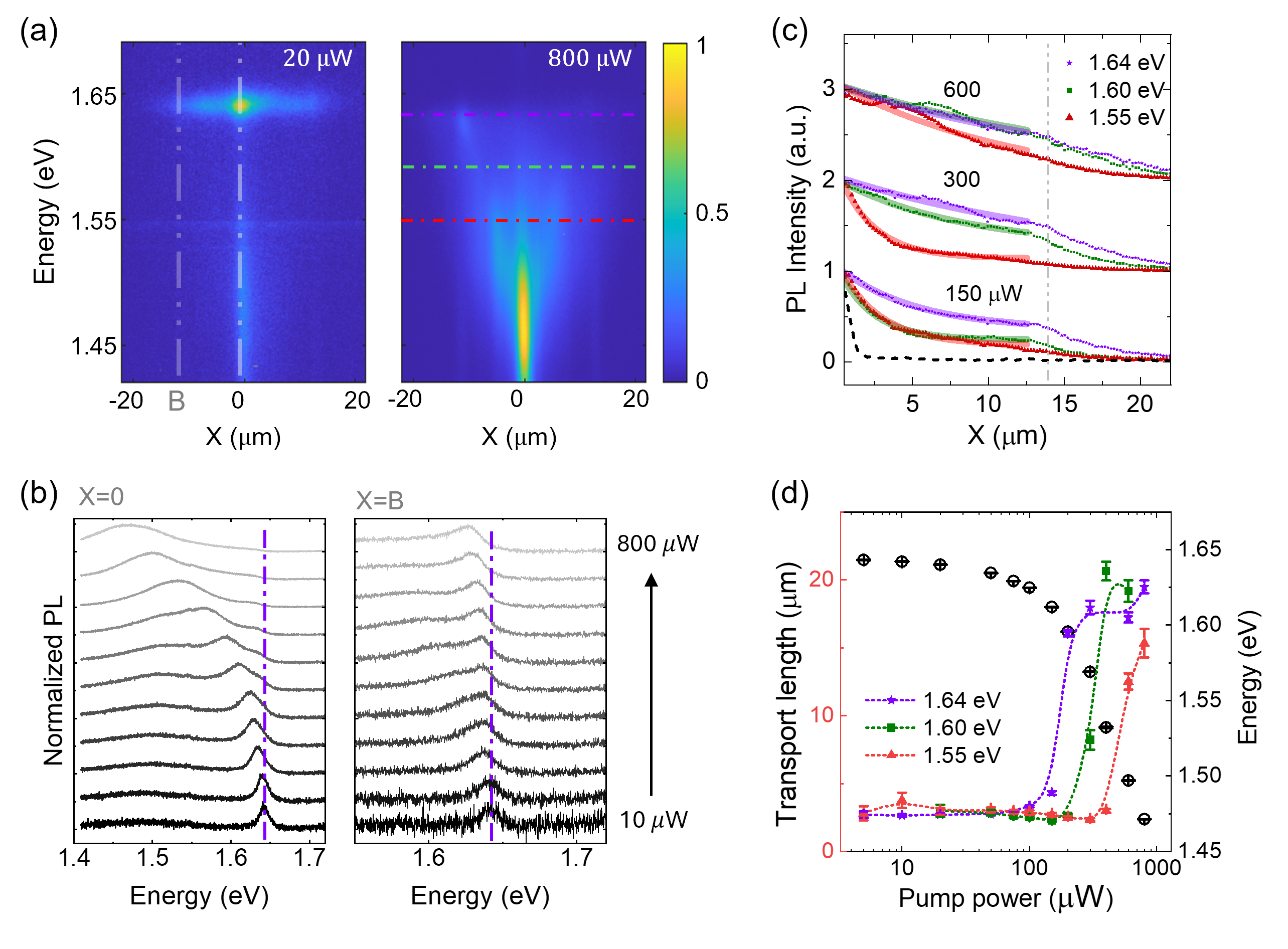}
\caption{\label{f3} Pump power dependence of the transport and energy relaxation properties.
\textbf{(a)} Spatially resolved PL spectra from PhC device c1 from pump powers of $P=20~\mu$W (left) and $800~\mu$W (right). \textbf{(b)} Pump-power dependence of normalized PL spectra at the excitation spot (left, $X=0$) and PhC boundary (right, $X=B$), indicated by the vertical lines in (a). The Purple lines mark the emission energy of $1.64$ eV. \textbf{(c)} Comparison of the pump power dependence of the spatial profiles of PL emission at different emission energies 1.64 eV (purple stars), 1.60 eV (green squares) and 1.55 eV (red triangles), as marked by the horizontal lines in (a). Profiles of different pump powers are offset along y-axis for clarity, from bottom to top, $P=150~\mu$W, $300~\mu$W, and $600~\mu$W. The PhC boundary is marked by the gray dashed line. The thick solid lines are single-exponential fits of the profiles inside the PhC. The black dashed curve is the exciton PL from the flat Si$_3$N$_4$ substrate. \textbf{(d)} Pump power dependence of the peak emission energy obtained from (b) (open black circles, right axis) and the transport lengths obtained from the fits in (c) (filled symbols, left axis). The transport lengths for different emission energies, 1.64 eV (purple stars), 1.60 eV (green squares) and 1.55 eV (red triangles), show different threshold powers of stimulated relaxation. 
}
\end{figure}

The above results reveal how the transport properties are correlated with the polariton band structure and polariton relaxation. Below we focus on PhC c1 to further study the effects of polariton relaxation by its pump power dependence. 
Figure~\ref{f3}(a) shows a comparison of the spectrally resolved spatial profile of PL from PhC c1 at low and high pump powers. At a low pump power, only the emission of exciton-like polaritons close to the exciton energy shows enhanced transport. At a high pump power, stronger emission is measured from lower energy states, which also shows increased transport length, suggesting increased scattering to lower-energy polaritons. 

These trends are shown more systematically in the pump power dependence of the spectra and spatial profiles of the emission in Figs.~\ref{f3}(b) and (c). 
Figure~\ref{f3}(b) shows the PL spectra at the excitation spot (left panel) and at the PhC boundary (right panel) as marked by the vertical lines at X = 0 and X = B in Fig.~\ref{f3}(a). At both positions, the emission redshifts with increasing pump power, suggesting more efficient scattering to lower-energy states at higher excitation densities. Lesser redshift at the PhC boundary suggests that higher energy states have a larger population and longer transport length.
Considering polaritons at three different energies (marked by horizontal lines in Fig.~\ref{f3}(a)), we compare in Fig.~\ref{f3}(c) how their spatial profiles change with excitation density. While they all show much longer transport lengths than the exciton emission, at low power, the highest energy emission at 1.64~eV has a significantly longer transport length than the lower-energy ones. With increasing power, a similarly long transport length is observed in lower-energy states.  


We summarize in Fig.~\ref{f3}(d) the pump power dependence of the emission energy, obtained from Fig.~\ref{f3}(b), and the transport lengths for different polariton states, obtained from Fig.~\ref{f3}(c) by an exponential fit. Beside the trends seen in Figs.~\ref{f3}(b)-(c), surprisingly, a highly nonlinear, threshold behavior becomes evident in both the emission energy and transport length. Below the pump power of $P=100~\mu$W,
the emission energy is close to the exciton energy and decreases slightly with increasing $P$; the transport length remains at merely about $2.6~\mu$m, likely dominated by exciton-like polaritons and localized states. Above $P\sim105~\mu$W, the emission energy redshifts significantly and continuously. Such a nonlinear change suggests that stimulated relaxation starts to dominate over spontaneous scattering into lower-energy polariton states. At the same time, a sharp jump in the transport length of the emission is observed, from $\sim2.5~\mu$m to $\sim20~\mu$m, first in high-energy polariton states at the pump power around $105~\mu$W, then at a higher threshold pump power for lower-energy polariton states. For each polariton state, the threshold pump power correlates with the redshift of the emission to that energy. Once above the threshold, the transport length stays around $20~\mu$m.

The above pump-power dependence consistently shows that the enhanced transport results from polariton states and depends on the polariton population distribution. The population distribution in turn depends on the exciton and polariton relaxation processes, controllable by the dispersion and excitation power separately. 
At low pump power, most emission comes from polariton states with high excitonic fractions and low group velocities, leading to a limited transport length. As the excitation density increases, the distribution shifts to lower-energy states with higher photonic fractions, larger group velocities, and longer transport. 
The shift is a result of stimulated relaxation of polariton, due to the bosonic nature of polaritons, manifested as the sudden increase of the transport length at the stimulated scattering threshold in Fig. \ref{f3}(d). 


Notably, the phenomena reported above are not limited to MoSe$_2$ at low temperatures, but also extends to other TMD materials, such as MoS$_2$, WSe$_2$, and WS$_2$ 
(see Supplementary Materials Fig. S2), and they persists even at room temperature (see Supplementary Materials Fig. S3), indicating their robustness and potential for device applications.

\section{Conclusion}
In short, we demonstrate enhanced and tunable transport in strongly coupled TMD-2D PhCs polariton devices. The formation of polaritons leads to an order-of-magnitude enhancement of the transport length of excitations compared to that of bare excitons. By varying the PhC parameters and controlling the excitation intensity, we change the transport behavior over a wide spectral range and length scales. Furthermore, pronounced stimulated relaxation is observed, which suggests the presence of nonlinear scattering processes and the potential to form polariton superfluids with frictionless transport. 
Such designable 2D material-2D PhC polariton systems may provide a versatile platform to implement novel transport properties and device functions.

\section*{Methods}

\textbf{Fabrication of 2D PhC.} The samples consist of a 140~nm thick Si$_3$N$_4$ layer on top of a Si substrate. The Si$_3$N$_4$ is deposited using the Low Pressure Chemical Vapor Deposition (LPhCVD) technique. The 2D PhC pattern was written into ZEP 520A E-beam resist through Electron Beam Lithography (JEOL JBX-6300FS). Subsequently, Inductively Coupled Plasma Reactive Ion Etching (ICP/RIE) was employed to etch the holes into the Si$_3$N$_4$ layer. After that, a gas etching using XeF$_2$ was applied to remove the underlying Si substrate, forming a suspended structure. Finally, the residual resist was removed using oxygen plasma etching.

\textbf{Large-area material preparation.} Dodecanol-encapsulated large-area monolayer MoSe$_2$ was prepared using the gold-tape exfoliation technique, following a similar procedure as previously reported\cite{Li_Macroscopic_2023}. Initially, a 150-nm gold film was deposited onto a Si wafer using E-beam evaporation. Subsequently, the wafer was spin-coated with a polyvinylpyrrolidone (PVP) solution. A single-sided heat release tape was then carefully attached to the PVP/gold surface, enabling the peeling off of the gold film. The obtained gold film was gently pressed onto a single crystal of TMD to exfoliate a monolayer of MoSe$_2$. Finally, the monolayer of MoSe$_2$ was transferred onto PhC structure with dodecanol self-assembled monolayer on top, followed by additional drop-cast of dodecanol layers on top of MoSe$_2$ for encapsulation to enhance the optical quality of MoSe$_2$.

\textbf{Optical measurements} The device was cooled down to 5~K using a Montana Fusion system. Real-space and momentum-space imaging of the device were performed using RC and PL measurements with a confocal microscopy system. An objective lens with a numerical aperture (NA) of 0.55 was used for both pumping and collection. For the RC measurement, a tungsten halogen lamp was utilized, providing a beam size of approximately 23 $\mu m$ in diameter. The RC was extracted by the ratio between the reflection on PhC and on uniform Si$_3$N$_4$. For PL measurements, a non-resonant excitation was performed using a continuous-wave solid-state laser operating at 532 nm with a beam size around 1 $\mu m$. To obtain the 2D spatial image of the exciton, a bandpass filter was used to filter the PL signal, and the signal was measured using a 2D array of detectors.
\newline

\noindent \textbf{Acknowledgement} \\
We acknowledge the support by the Army Research Office under Awards W911NF-17-1-0312, the Air Force Office of Scientific Research under Awards FA2386-21-1-4066, the National Science Foundation under Awards DMR 2132470, the Office of Naval Research under Awards N00014-21-1-2770, and the Gordon and Betty Moore Foundation under Grant GBMF10694. 

\noindent \textbf{Data availability}\\
Source data will be deposited into a publicly available repository. 
%

\noindent \textbf{Competing interests}\\
The authors declare no competing interests.

\noindent \textbf{Author contributions} \\
X.X. and H.D. conceived and designed the research.
X.X. performed the simulation, fabrication, measurements, and data analysis. 
Q.L. conducted the large-scale preparation and transfer of materials and assisted in measurements. 
C.L. (Chenxi L.), Y.L. and C.L. (Chulwon L.) assisted in fabrication.
K.S. and H.D. assisted in data analysis. 
X.X. and H.D. wrote the manuscript. 
All authors read and commented on the manuscript.\\

\bibliographystyle{unsrt}
\bibliography{main}


\end{document}